\documentclass[12pt]{article}

\usepackage{newtxtext,newtxmath}

\usepackage{graphicx}

\usepackage[letterpaper,margin=1in]{geometry}

\def\be{\begin{equation}} 
\def\ee{\end{equation}}
\newcommand \bea {\begin{eqnarray}} 
\newcommand \eea {\end{eqnarray}}

\renewenvironment{abstract}
	{\quotation}
	{\endquotation}

\date{}

\makeatletter
\renewcommand{\fnum@figure}{\textbf{Figure \thefigure}}
\renewcommand{\fnum@table}{\textbf{Table \thetable}}
\makeatother

\usepackage{scicite}

\usepackage{url}

\def\scititle{What do the fundamental constants of physics tell us about life?}
\title{\bfseries \boldmath \scititle}

\author{
	Pankaj Mehta$^{1\ast}$ 
	Jan\'e Kondev$^{2\ast}$\and
	\small$^{1}$Department of Physics and Faculty for Computing and Data Science, Boston University, Boston MA \& 02215, USA.\and	
	\small$^{2}$ Martin A. Fisher School of Physics, Brandeis University, Waltham MA \& 02453, USA.\and
	\small$^\ast$Corresponding authors. Email: pankajm@bu.edu; kondev@brandeis.edu }

\begin{document} 

\maketitle

\begin{abstract} \bfseries \boldmath
In the 1970s, the renowned physicist Victor Weisskopf famously developed a research program to qualitatively explain properties of matter in terms of  the fundamental constants of physics. But there was one type of matter prominently missing from Weisskopf's analysis: life. Here, we develop Weisskopf-style arguments demonstrating how the fundamental constants of physics can be used to understand the properties of living systems. By combining biophysical arguments and dimensional analysis, we show that vital properties of chemical self-replicators, such as growth yield, minimum doubling time, and minimum power consumption in dormancy, can be quantitatively estimated using fundamental physical constants. The calculations highlight how the laws of physics constrain chemistry-based life on Earth, and if it exists, elsewhere in our universe.

--------------------------------------
\end{abstract}

\noindent
A defining achievement of twentieth century physics is the realization that the properties of matter can be understood using the fundamental theories of physics. Features as diverse as the density of rocks, the visibility of compact matter, and the size of mountains, all emerge from quantum mechanical  interaction between electrons, nuclei, and electromagnetic radiation. Motivated by this observation, more than fifty years ago, the physicist Victor Weisskopf developed an influential research program showing that it was possible to semi-quantitatively understand the properties of terrestrial matter in terms of six fundamental constants: the speed of light $c$,  Planck's constant $\hbar$ (the constant governing quantum fluctuations), the charge of an electron $e$, the mass of the electron $m_e$, the mass of the proton $m_p$, and Newton's gravitational constant $G_N$  \cite{weisskopf1970physics, weisskopf1975atoms,burrows2014astronomical}. For example, using clever physical argumentation Weisskopf famously provided order-of-magnitude estimates for the maximum height of mountains on Earth ($\sim 26$ km) and Mars ($\sim 50$ km) \cite{weisskopf1970physics, weisskopf1975atoms}. Weisskopf argued that such qualitative arguments complement more rigorous quantitative calculations by highlighting the essential physics needed to explain a phenomenon.

\begin{figure} 
	\centering
	\includegraphics[width=0.6\textwidth]{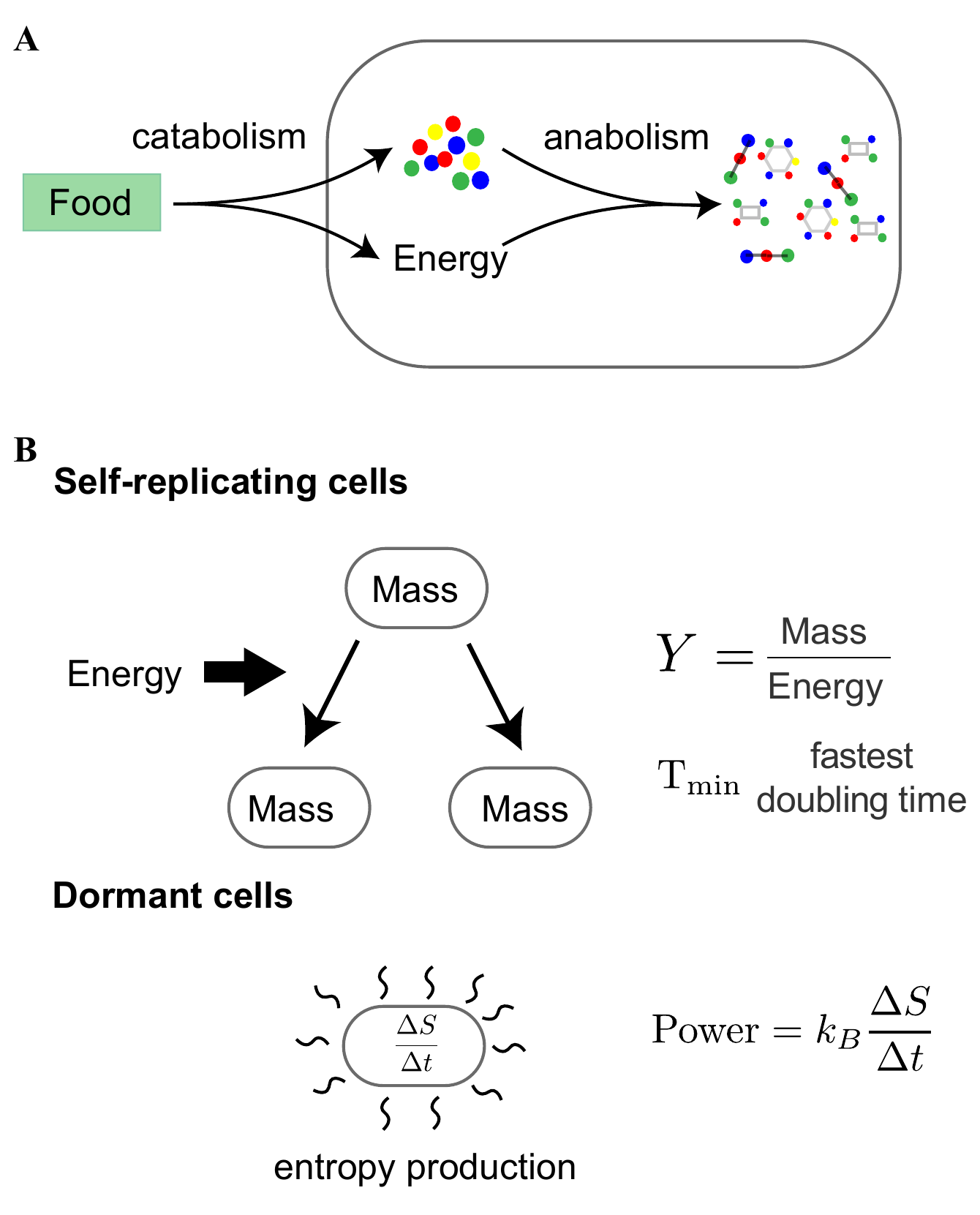} 

	\caption{\textbf{	Characterizing chemical self-replicators.} (a) Self-replication requires	organisms to break down food molecules from the environment to extract energy and metabolic precursors (catabolism) and then use this energy to synthesize the complex molecules necessary for self-replication (anabolism). (b) We focus on three properties of self-replication: (i) the mass per unit energy consumed needed to self-replicate, (ii) the time it takes for self-replication, and (iii) the energy per unit time a dormant cell must expend in order to stay alive and counter entropic forces.}
\label{figure1} 
\end{figure}

Weisskopf applied his program of explaining the world in terms of fundamental physical constants to all kinds of matter ranging from rocks, to water waves, to stars. However, there was one prominent form of matter missing from Weisskopf's original analysis: living systems. From a physical perspective, life is a novel form of non-equilibrium self-organized matter whose defining feature is high-fidelity self-replication \cite{schrodinger1992life}. This process requires living organisms to carry out two general classes of metabolic tasks: extracting energy from the environment (catabolism) and synthesizing the new material needed for self-replication (anabolism) \cite{scott2010interdependence, scott2023shaping,spormann2023principles} (see Figure \ref{figure1}).

	
In chemistry-based life like that found on Earth, self-replication is powered by transferring electrons from a  high-energy state to a lower-energy final state. For example, in aerobic respiration electrons in high-energy molecules like sugars are transferred to low-energy oxygen molecules. This observation led Nobel-prize winning biochemist Albert Szent-Gy\"orgyi to quip that "Life is nothing but a high energy electron looking for a place to rest'' \cite{trefil2009origin, szent1968bioelectronics}. This reduction of life to electrons of course misses much of what makes living systems so interesting -- the central role played by information transmission, ecology, evolution, and complexity \cite{fisher2013evolutionary, hartwell1999molecular, kirschner2000molecular}. Nonetheless, it has the virtue of highlighting the fact that the law of physics, and in particular quantum mechanics and thermodynamics, still fundamentally constrain life. This suggests it should be possible to extend Weisskopf-style argumentation to understand living matter \cite{dill2011physical}.

In order to carry out such a program, we draw on the rich emerging literature quantifying biological phenomena across living systems \cite{moran2010snapshot, milo2010bionumbers, milo2015cell}. For concreteness, we focus here on three properties of chemical self-replicators: (1) the growth yield $Y$ which measures the amount of biomass produced (in grams) per amount of energy consumed; (2) the minimum time for self-replication; and (3) the minimum power consumption $P_{min}$ (J/s) needed by a dormant cell in order not to die (see Figure \ref{figure1}). 

For example, as discussed by \cite{spormann2023principles}, for a wide variety of single-cell organisms the growth yield in energy-limited conditions is about $Y \sim1 \times 10^{-4}$\,g J$^{-1}$. This number is often reported as  $Y \sim 5-10$\,g (mol ATP.)$^{-1}$. To convert between moles ATP and Joules, we have used the fact that 1 mol ATP yields  $\sim 50$\,kJ in standard conditions. Similarly, a mix of theoretical arguments and empirical measurements suggest that the fastest self-replication time for a bacterial cell is approximately $\tau_{double} \approx 5 \times 10^2$\,s \cite{dill2011physical, milo2010bionumbers}.  In contrast, bacteria in extreme environments have doubling times that are almost eight orders of magnitude larger \cite{martens2009ammonia, orcutt2011microbial, lloyd2020evidence}. Can we understand the origin of these numbers in terms of fundamental physical constants?  Recent lab experiments suggest that the minimum maintenance power of dormant bacterial cells is approximately $10^3$ ATP/s/cell, which translates to a $P_{min} \approx 10^{-16}$\,J/s/cell \cite{ciemniecki2024mechanistic}, consistent with earlier empirical estimates of $P_{min}$ based on measurements of metabolic rates of microbial communities in environmentally derived samples \cite{price2004temperature}. How do the laws of physics give rise to this?

\section*{Physical scales and fundamental constants}

Our starting point for understanding these numbers is the fact that the physical processes underlying both catabolism and anabolism in chemical self-replicators are governed by quantum mechanical interactions between electrons and nuclei. These interactions are associated with a characteristic energy scale, the Rydberg energy $Ry \approx 13.6$ eV $\approx 2.18 \times 10^{-18}$ J, and a characteristic length scale, the Bohr radius  $a_0 \approx 0.53 \mathring {\mathrm A}  \approx 5.3 \times 10^{-11}$~m. Weisskopf gave a simple qualitative argument for estimating these scales using energy minimization and the de Broglie relation, $p=\hbar/\lambda$, which relates the momentum of a particle, $p$, to its wavelength, $\lambda$, through Planck's constant $\hbar= 1.05 \times 10^{-34}$ J$\cdot$s \cite{weisskopf1975atoms}. 

The energy of an electron with momentum $p$ confined within a atom of radius $r$ can be written as a sum of the electron's kinetic energy and potential energy due to electromagnetic interactions $E= {p^2 \over 2  m_e }-{1 \over 4 \pi \epsilon_0} {e^2 \over r}$, where $\epsilon_0$ is the vacuum electric permittivity, $e$ is the electron charge, and $m_e$ is the mass of the electron. Since a particle confined to a region of size $r$ will have a wavelength $\lambda \sim r$, we can use the de Broglie relation to write the characteristic momentum associated with a confined electron, namely $p \sim \hbar /r$. Substituting this into the expression above yields the energy of the electron as a function of the confinement distance,
\be
E(r) \approx {\hbar^2 \over 2 m_e r^2} -{1 \over 4 \pi \epsilon_0} {e^2 \over r}.
\label{E:E(r)}
\ee
The electron is typically located at a distance that minimizes its total energy, allowing us to calculate the Bohr radius $a_0$ by finding the value of $r$ that minimizes $E(r)$.  A straightforward calculations yields (see Methods)
\be
a_0={4 \pi   \epsilon_0 \hbar^2 \over m_e e^2}= {\hbar \over \alpha m_e c},
\label{E:a0}
\ee
where in going to the second expression we have introduced the fine structure constant $\alpha \approx 1/137$, the speed of light, $c$, and used definition of the vacuum electric  permittivity $\epsilon_0 = {e^2 \over 4\pi \alpha \hbar c}$. Substituting $a_0$ into Eq.~\ref{E:E(r)} gives us the typical energy associated with an electron at a distance $a_0$, the Rydberg energy 
\be
Ry=-E(a_0)={e^2 \over 8 \pi \epsilon_0 a_0}= {1 \over 2}{\alpha^2 m_e c^2}.
\label{E:Rydef}
\ee
 The Rydberg energy, $Ry$, and the Bohr radius, $a_0$, are the fundamental length and energy scales associated with chemical physics. 

We can also use the Rydberg energy to estimate the energy, $E_{bond}$, associated with breaking a chemical bond. To do so, we exploit the empirical observation that the size of a chemical bond $l_{bond}$ is generally a few Bohr radii so that we can write $l_{bond} \sim f_{bond} \,a_0$. We adopt the notation that singles out the dimension-full quantity ($a_0$) that sets the order of magnitude of the quantity of interest and clump the rest into a numerical factor, $f_{bond}$. For example, for a carbon-carbon bond, $f_{bond} \approx 3$. Since the chemical bond corresponds to "shared" electrons between atoms, we can estimate its energy by the Coulomb energy of an electron whose distance to the positively charged nucleus is approximately equal to the bond length (see Eq. ~\ref{E:Rydef}). From this simple physical picture we can immediately conclude that the bond energy scales inversely with the bond length, i.e., $E_{bond} \sim Ry/f_{bond}$. For $f_{bond}=3$, this expression yields $E_{bond} \approx 4.5$ eV which is comparable to the empirically measured energy to break a C-C bond, $350$ kJ/mol or $3.6$ eV/bond.

The physics of life is not governed solely by quantum mechanics. Instead, it emerges from a delicate interplay between thermal and quantum fluctuations \cite{phillips2006biological}. The reason for this is that self-replication is an inherently non-equilibrium phenomenon. The physical constant governing the scale of thermal fluctuations is the Boltzmann constant, $k_B=1.4 \times 10^{-23}$ J$\cdot$K$^{-1}$.
We can characterize the importance of thermal fluctuations at a temperature $T$ in terms of the thermal energy, $k_B T$. At standard temperatures relevant to life on earth ( $T=298K$), $k_B T = 4  \times 10^{-21}$ J or in more relevant units for us, $25.7$ meV. Notice that the thermal energy scale $k_B T$ is more than 500 times smaller than the Rydberg energy, Ry, and about 100 times smaller than $E_{bond}$. This vast difference in energy scales between quantum mechanics and thermal energy underlies the extreme stability of chemical bonds and  explains why chemical reactions must be actively catalyzed in living systems.

Thus, far we have focused on identifying relevant energy and length scales. However, self-replication is a dynamics process that also requires us to understand the time scales of chemical processes. The reaction rates of chemical reactions are limited by the ability of molecules to find each other through diffusion in their fluid environment.  In the diffusion limited regime, a molecule of size $r$ at a concentration $c$ encounters a target of size $R$ at a rate $k_{on}=4 \pi D c R$, where $D$ is the diffusion constant \cite{berg1977physics}. We show below that there is a maximum rate $k_{on}^{max}$ at which molecules can find a target. For a molecule diffusing in a fluid of concentration $c_f$ with atomic number $A_f$ and temperature $T$, we have 
\be
k_{on}^{max}= A_{f}^{-{1 \over 2}}\left({4 \pi \over 3} \right) \times  \left({c \over c_f}\right) \times \left(R \over r \right) \times \tau_{min}^{-1}(T)
\label{E:kon}
\ee
where we have defined the characteristic ``kinetic time''
\be
\tau_{min}(T)={\hbar \over k_B T}\left( { m_e \over m_p}\right)^{-{1 \over 2}}.
\label{E:tau_min}
\ee
 At $37^\circ$ C, $\tau_{min}$ is about $1$ picosecond. For molecules at a concentration of 100mM in water, like amino acids in the cytoplasm, the numerical value of $k_{on}^{max}=2 \times 10^9$ s$^{-1}$. In calculating this, we have used the fact that for water $A_f=18$ and $c_f =55$M.
 Amusingly, the dominant numerical contribution to $\tau_{min}$ comes from the ratio ${\hbar / k_BT}$, highlighting how chemical kinetics emerges from a subtle interplay between thermal and quantum fluctuations. As expected, for higher temperatures thermal fluctuations get stronger and this time scale gets smaller.

A full derivation of the expression in Eq.~\ref{E:kon} is given in the Methods. Here, we briefly outline the physical intuition underlying this expression. Our starting point is the observation that we can re-express the diffusion constant in terms of the temperature $T$
and the fluid viscosity $\eta$ using the Stokes-Einstein relation $D= {k_B T \over 6 \pi \eta r}$. Rather than work with the viscosity directly, it is useful to rewrite this relation in terms of the kinematic viscosity, $\nu=\eta/\rho_f$, where $\rho_f=A_f m_p c_f$ is the mass density of the fluid. In general, both $\eta$ and $\nu$ are a function of temperature. However, Ed Purcell pointed out that varied fluids all exhibit a similar numerical value for the minimum kinematic viscosity $\nu_{min}$ \cite{purcell2014life, trachenko2021quantum}. Surprisingly, $\nu_{min}$ does not depend on temperature and is entirely a property of quantum mechanical interactions between fluid molecules. The natural scale for $\nu_{min}$ is set by a combination of fundamental constants we call the Berg viscosity, $\nu_B ={1 \over 2 \pi} {\hbar \over \sqrt{m_e m_p}}$, with $\nu_{min}= A_f^{-1/2} \nu_B$ (\cite{trachenko2021quantum} and Methods). Since $k_{on}$ is inversely proportional to  $\nu$, substituting $\nu_{min}$ into the expression for $k_{on}$ allows us to calculate the fastest possible kinetic rate, $k_{on}^{max}$, resulting in Eq.~\ref{E:kon}. Finally, we note that we can express $\tau_{min}(T)$ in terms of the the Berg viscosity as $\tau_{min}(T)/(2 \pi)=m_p \nu_{B}/k_B T$.

\begin{figure} 
	\centering
	\includegraphics[width=0.6\textwidth]{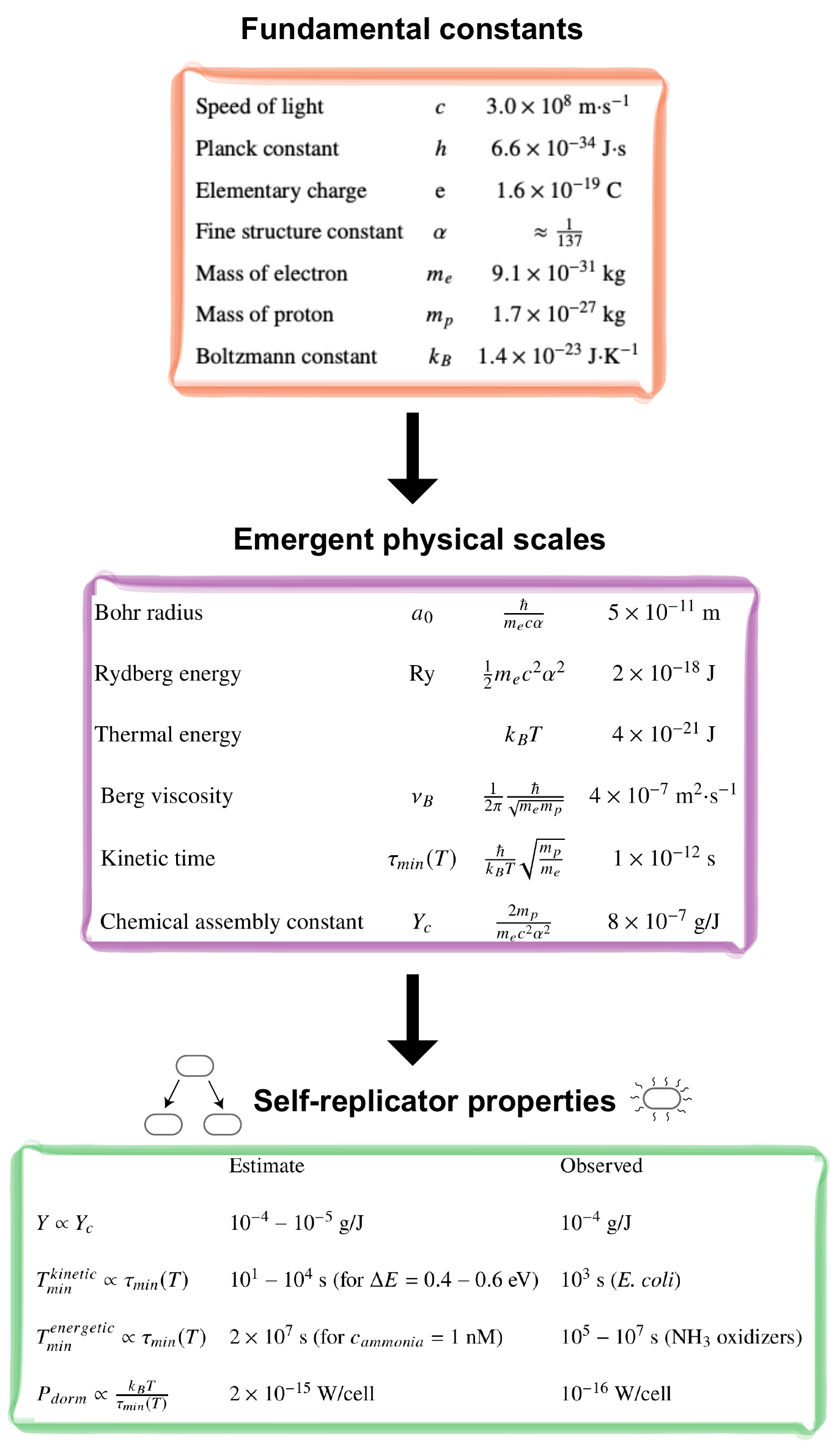} 

	\caption{\textbf{	From fundamental constants to self-replicator properties.} The fundamental physical constants give rise to emergent physical scales that govern the properties of chemical self-replicators.}
	\label{fig:fig2} 
\end{figure}

\section*{Properties of chemical self-replicators}

The relevant fundamental physical constants and emergent physical scales are summarized in Figure \ref{fig:fig2}. We now show how these quantities can be used to quantitatively estimate the properties of living systems

\subsection*{Mass yield} 

A defining property of life is self-replication. This requires organisms to extract energy from their environments in order to create new biomass. One interesting quantity that we can use to  characterize this process is the mass yield, $Y$, which measures the amount of new mass produced by the organism per unit of energy consumed. As discussed above,  for a wide variety of single-cell organisms, experimental measurements suggest that $Y \sim 1\times10^{-4}$\,g J$^{-1}$. How can we understand this quantity in terms of fundamental physical scales discussed above?

Our starting point is observation that the dominant source of energy consumption during self-reproduction are anabolic processes that synthesize new biomass. Consider an organism of dry mass $M$ that requires the synthesis of $N_{bond}$ bonds at an energetic cost $E_{bond}= Ry/f_{bond}$. If the synthesis of these new bonds is the dominant source of energy consumption, the mass yield can be written as $Y= M/(N_{bond} E_{bond})= f_{bond}  M /( N_{bond} Ry )$. Generically, the number of new bonds that need to be formed will be proportional to the number of atoms $N$ in the organism, so that $N_{bond} \approx  b N$, with $b$ a numeric proportionality constant. The number of atoms $N$ and the mass $M$ are related through the average molecular weight of the biomolecules, namely, $M=NA_{bio}m_p$ where $m_p$ is the mass of a proton and $A_{bio}$ is the average atomic mass of constituent biomass molecules. Rearranging and re-expressing the Rydberg energy $Ry$ in terms of fundamental constants, yield the expression
\begin{align}
Y &\approx \left({A_{bio} f_{bond} \over b}\right) \times Y_c,
\end{align}
with 
\be
Y_c={2 m_p \over m_e c^2 \alpha^2}.
\ee
$Y_c$ defines a new physical scale we call the chemical assembly constant with a numerical value $\sim 8 \times 10^{-7}$ g/J. A striking property of this expression is that it depends only on the composition of an organism and not on its size or mass. 

We can compare the predictions of this expression to experimental observations in archaea and bacteria. To do so, we makes use of the fact that life on earth is composed largely of carbon so that $A_{bio} \approx 13$ and $f_{bond} \approx 3$. We can also estimate $b$ by exploiting the fact that the bulk of newly synthesized biomass are proteins. Proteins are composed of amino acids, which range in size from 10-30 atoms. Depending on the medium in which cells are grown, amino acid synthesis can require anywhere from 0 bonds (i.e. all amino acids are provided in the medium) to 15-25 bonds in rich medium. In addition, making proteins requires synthesizing one peptide bond per amino acid. This means that $b \sim1$ in minimal media and $b \sim1/10$ in rich media. For these choices, we have that $Y \approx 3 \times 10^{-5}$ g/J and $Y \approx 3 \times 10^{-4}$ g/J, respectively. These estimates compare favorably to the experimentally observed values of $1 \times 10^{-4} g/J$.

\subsection*{Time for replication} 
We now change our focus from energetics to dynamics. Single-celled organisms exhibit an incredibly wide range of doubling times. {\it E. coli} in rich media doubles in 20 minutes, whereas ammonia-oxidizing archaea take a day to several months to self-replicate, with the exact time depending on environmental conditions \cite{martens2009ammonia}. Lithoautotrophic bacteria in the deep sea crust have even longer doubling times, estimated to be of order years \cite{lloyd2020evidence}. To understand the physical origins of these numbers, it is helpful to distinguish between two possible scenarios. The first is that growth is kinetically limited. In this case, the cell division time is limited by the time it takes to carry out the chemical reactions  needed to synthesize new biomass \cite{dill2011physical,tlusty2025life}. The second possibility is that self-replication is energy-limited. In this case, the division time is set by the time it takes to extract sufficient energy from the environment to create a new organism \cite{spormann2023principles}.

For kinetically limited replicators, it is essential to estimate the time scale associated with chemical reactions. To do so, we will make use of  transition-state theory in chemical kinetics. Within this framework, the rate of a chemical reaction is simply the product of the diffusion limited on rate, $k_{on}^{max}$, and an Arrhenius factor, $e^{-\Delta E \over k_B T}$, stemming from the fact that the reactions cannot happen spontaneously and must pass through a high-energy intermediate  transition state. If the transition state has energy $\Delta E$, then 
\be
k^{max}=k_{on}^{max} e^{-\Delta E \over k_B T},
\label{E:kmax}
\ee
 with $k_{on}^{max}$ given by Eq.~\ref{E:kon}.

To calculate $T_{min}$, we need an estimate of the activation energy $\Delta E$  of the transition state. One qualitative way of estimating $\Delta E$ is to model chemical bonds as springs and ask how much bonds are stretched or compressed at the transition state (see Methods). The  ``spring constant'' of a bond of length $l_{bond}=f_{bond}a_0$ can be estimated from the bond energy using Hooke's law,  $E_{bond}=K_{bond} l_{bond}^2/2$. Using the fact that $E_{bond}=Ry/f_{bond}$, we can write the spring constant as function of the Rydberg energy as $K_{bond} =2 Ry/(f_{bond} l_{bond}^2)$. At the transition state, the bond length changes by $\Delta l$. Let us denote the fractional change in the length of the bond by $\delta=\Delta l/l_{bond}$. The activation barrier of the transition state is $ \Delta E^{min} = {1 \over 2}K_{bond} (\Delta l)^2 = \delta^2 Ry/f_{bond}$ . For $f_{bond} =3$ and $\delta =0.3-0.5$ -- a reasonable choice for a wide variety of reactions -- one gets that  $\Delta E \approx 0.4-1.1 \mathrm{eV}$. This is comparable to the observed activation energy for most common reactions.

We are now in a position to calculate the minimum doubling time $T_{min}$ for kinetically limited self-replicator. The fastest self-replicator is an organism where each molecule replicates itself  \cite{dill2011physical}. If a molecule on average has $N_{M}$ bonds that need to be synthesized, then the  minimum doubling will simply be
\be
T_{min}=N_{M}/k^{max}=N_M  \times A_{f}^{{1 \over 2}}\left({3 \over 4 \pi} \right) \times \left({c_f \over c}\right) \times e^{ {\delta^2 \over  f_{bond}}{Ry \over k_B T}} \times  \tau_{min}(T),
\ee
where $c_f$ is the fluid concentration and $c$ is the concentration of precursor molecular building blocks. In writing this, we have used the fact that the molecules are of roughly same size so that $R \approx r$ in Eq.~\ref{E:kon}.

For  {\it E. coli} growing in exponential growth at $37^\circ$ C, the biomass production of the cell is dominated by the production of new ribosomes  which account for the bulk of the dry mass \cite{scott2023shaping}. A good approximation to this situation is to assume that every protein translates itself \cite{dill2011physical}. In this case, we can choose $N$ to be equal to the typical number of peptide bonds in all ribosomal proteins, $N_M \sim 7400$. The typical concentration $c$ of amino acid precursors is $100$ mM resulting in an $k_{on}^{max}=2 \times 10^9$ s$^{-1}$. For $\Delta E$ in the range $0.4-1.1 \mathrm{eV}$, this give $k^{max}$ in the range $5 \times 10^2 \mathrm{\,s}^{-1}$ to $3 \times 10^{-9} \mathrm{\,s}^{-1}$, reflecting the extreme sensitivity of this rate to the activation barrier. Using these rates, one finds that $T_{min} \approx 10s$ for $\Delta E=0.4$ eV,  $T_{min} \approx 480s$ for $\Delta E=0.5$ eV, and $T_{min} \approx 10^{11}$s for $\Delta E=1$ eV. 

These numbers can be compared to experiments. The fastest known translation rate is about 20 amino acids per second, corresponding to a $k^{max}=20 \mathrm{\,s}^{-1}$. The fastest observed doubling time is for the bacterium {\it Vibrio natriegens} , which has a  $T_{min} \approx 600$ s. While our estimates span a wide range,  they are consistent with these experimental observations and put a strong lower bound on the fastest doubling times. However, the exponential dependence of chemical kinetic constants on activation energy make it difficult to make more stringent predictions for kinetically-limited self-replicators using purely qualitative methods.

In energy-poor environments, self-replication is not limited by kinetics, but by the ability of organism to harvest electron donor molecules from the environment.  If we denote the rate at which bacteria can extract energy from the environment by $P_{E}$, then we can qualitatively estimate the division time for energy limited replicators using the formula $T_{min}=E_{tot}/P_E$, where $E_{tot}=N_{bond}E_{bond}$ is the total energy needed for an organism to copy itself. In writing this last expression, we have once again assumed that the dominant energy expenditure is the synthesis of new bonds. It is helpful to once again express $N_{bond}$ in terms of the mass of an organism, namely $N_{bond}= { b M \over A_{bio} m_p}$, with $b$ the average number of bonds per atom.

In order to estimate $P_E$, we model bacteria as absorbing spheres of size $R \approx 1$ micron that capture electron donor molecules found in the environment at a concentration $c$ of size $r \approx 1$ nm. The rate at which bacteria can extract energy is limited by the diffusion limited flux incident on the cell, which can be calculated from Eq.~\ref{E:kon}.  For ammonia-oxidizing archaea, $c$ is typically in the nM range whereas for deep sea chemolithoautotrophs, the limiting source of energy is dissolved $H_2$ molecules which are found at  sub-nanomolar concentrations, $c \approx 10^{-10}-10^{-12}$ M. If organisms can extract an energy $E_{redox}$ per molecule, the power harvested from the environment is $P^{max}=k_{on}^{max} E_{redox}$. $E_{redox}$ is set by the details of redox chemistry and is typically of order $0.1-1 eV$.  Combining these expressions and using the fact $E_{bond}=Ry/f_{bond}$, one has
\be
T_{min} =A_{f}^{{1 \over 2}}\left({3 b \over 4 \pi A_{bio} f_{bond}} \right) \times \left({c_f \over c}\right) \times \left(r \over R \right) \times  \left({M \over m_p}\right) \times \left( {Ry \over E_{redox}}\right) \times  \tau_{min}(T)
\ee
Using a value of $M=10^{-12}$ g, $A_{bio} \approx 13$, $b \approx 1$, $E_{redox} \approx 0.77$ eV, and $T \approx 300K$, we arrive at an estimated doubling time of $T_{min}= 2 \times 10^{7}$s or about 6 months for ammonia oxidizers and $T_{min}= 2\times 10^8 - 2 \times 10^{10}$s or 5-500 years for bacteria that live deep in the ocean crust, consistent with experimental observations \cite{martens2009ammonia, orcutt2011microbial, lloyd2020evidence}.

\subsection*{Power consumption of dormant cells} 
Thus, far we have focused on the energetics and kinetics of reproduction. However, cells must expend energy even when they are not growing \cite{walker2024microbial}. The reason for this is that life is inherently a non-equilibrium phenomenon that requires energy in order to counteract entropic forces.  Paraphrasing Schr\"odinger, life feeds on ``negative entropy'' \cite{schrodinger1992life}. For compartmentalized organisms with a membrane, one such process arises from the need to maintain the cellular membrane potential $V_m$ resulting from concentration differences of charged ions inside and outside the cell \cite{schink2024survival,walker2024microbial}. Even though the membrane is an excellent insulator it still supports a "leakage" ion current across the membrane. This leads to loss of membrane potential which cells actively counteract by using energy-consuming ion pumps. 
Here, we assume that this process dominates the power consumption of dormant cells, $P_{dorm}$. Other processes such as maintaining the proteome in light of spontaneous protein degradation and counteracting the loss of post translational modifications also likely contribute to $P_{dorm}$ \cite{Salmonthesis}. However, estimates based on empirical measurements of environmental samples suggest that these processes may be up to three orders of magnitude less costly than maintaining the membrane potential \cite{price2004temperature}.

The simple idea we fully explore in the Methods, is that thermal fluctuations open pores in the membrane which allow ions to pass through. There we show that this requires the cell to consume energy at a rate  
\be
P_{dorm}={1 \over 3}A_{f}^{-{1 \over 2}} \times \left({c_i \over c_{f}}\right) \times n_{pore} \times \left({S_{p} \over  d_m r}\right) \times \left[\ln{ \left({c_{o} \over c_{i}}\right)}\right]^2 \times {k_B T \over  \tau_{min}(T)},
\label{E:Pmin}
\ee
where $n_{pore}$ is the expected number of pores formed due to thermal fluctuations, $S_p$ is the surface area of the membrane pore, which we take to have the  radius that is typical of a hydrated ion (i.e., a few \AA);  $d_m$ is the membrane thickness, $r$ is the size of an ion, and $c_{o/i}$ is the concentration of charged ions outside/inside the cell. Notice that the fundamental physical constants are all contained in the term, $k_B T /  \tau_{min}(T)$, which has a numerical value of $4 \times 10^{-9}$ J/s at room temperature. This suggests that natural physical scale characterizing entropy production in living system is simply thermal energy $k_B T$ divided by the kinetic time scale $\tau_{min}(T)$. 

We briefly outline the physical origins of this expression.  A full derivation is provided in the  Methods. As discussed above, the membrane voltage $V_m$ gives rise to a leak current $I$ across the membrane. From elementary physics, we know that the power dissipated by this current is just $P=IV_m$. Due to the second law of thermodynamics,  cell must expend at least this much energy per unit time to counteract leak currents. For this reason, the problem of calculating $P_{dorm}$ reduces to deriving expressions for  $V_m$ and the maximum possible leak current $I^{max}$.

The membrane potential can be calculated using standard thermodynamic arguments. The electrical potential energy due to $V_m$ must equal to the free energy resulting from concentration differences, resulting in the identity $ V_m = {k_B T \over z e}\ln{ \left({c_{o} \over c_{i}}\right)}$, where $z$ is the ion charge. The prefactor ${k_B T/  e}$ is often called the thermal voltage and is  approximately $26$ mV at room temperature ($T=300$ K) \cite{milo2015cell}. For a membrane of thickness $d_m$, this voltage difference gives rise to an electric field of strength $E_m=V_m/d_m$ across the membrane. The resulting force accelerates the ions to a velocity $v_i$ at which point the force due to $E_m$ is balanced by the frictional drag force on ions from the surrounding fluid, $zeE_m=\gamma v_i$,  with $\gamma$ the drag coefficient. The corresponding leak current is just the net charge that crosses the membrane per unit time, $I=ze (c_o-c_i) S_p v_i $. 

To calculate $v_i$, we use the fact that, at steady-state, the electrical and drag forces must balance, resulting in the relation $ze V_m/d_m=\gamma v_i$ \cite{NelsonEMP}. For ions of size $r$, we can relate $\gamma$ to the kinematic viscosity $\nu$ using the Stokes relation, $\gamma=6 \pi A_f c_f \nu r$.  Notice that the velocity, and hence the current, will be maximal, when $\gamma$, and hence the kinematic viscosity $\nu$ is at its minimum value. As discussed above, this happens when $\nu =\nu_{min}$. This observation allows us to calculate the maximum leak, resulting in Eq.~\ref{E:Pmin}.

We can use Eq.~\ref{E:Pmin} to estimate the power consumption of a single membrane pore. Then, by estimating the average number of such pores, $n_{pore}$, that are generated in the cell membrane by thermal fluctuations, we arrive at the power consumption of a dormant cell, $P_{dorm}$. We note that $c_i \approx 300$ mM,  $d_m \approx 4$nm ,  $r \approx 0.3$ nm,   $S_{p}\approx 0.3$ nm$^2$, and $\ln{ \left({c_{o} \over c_i}\right)} \approx 3$. Plugging in these numbers gives estimates for the maximal current, $I^{max} \approx 50 \times 10^{-12}$ A, and power dissipation $P_{max} \approx 4 \times 10^{-12}$ W for a single membrane pore. To estimate $P_{dorm}$, we simply multiply the power dissipated by a single pore by the expected number of pores ($n_{pore} \approx 5 \times 10^{-4}$). To make the estimate for $n_{pore}$ we make use of an estimate for the probability of making a pore ($\approx 7 \times 10^{-11}$), which we arrive at by estimating the energy cost of a pore ($\approx 0.6$eV), and by estimating the number of possible pore sites in the inner membrane of an {\em E.coli} cell ($\approx 7 \times 10^6$). This gives an estimate $P_{dorm}  \approx 2 \times 10^{-15}$ W/cell, or $2 \times 10^4$ ATP/sec/cell, where in going between these numbers we have used the fact that the free energy of ATP hydrolysis in cells is $\Delta G \approx 10^{-19}$ J. This is within a factor of $\sim 20$ of recent measurements that estimated the power consumption rate of a cell to be $\sim 10^3$ ATP/s, or equivalently $10^{-16}$ W \cite{ciemniecki2024mechanistic}. This is also similar to the observed values for dormant bacteria in natural environments \cite{price2004temperature}.

 This is remarkably good agreement given the fact that the probability of making a membrane pore is exponentially dependent on the energy cost of pore formation. This means that a small change in the estimate of the pore formation energy such as $(\ln10) k_B T = 0.059$~eV leads to an order of magnitude change in probability of pore formation and equally large change in maintenance cost. Interestingly, data on maintenance energy for bacteria isolated from  agricultural and forest soils, marine sediments, as well as polar ice cores, show a much larger variability than the energy requirements for growth, and span at least two orders of magnitude \cite{price2004temperature}. This we would expect based on pore formation energy depending slightly, even at the 10\% level, on the lipid composition of the cell membrane. 

\section*{Conclusion}

Life, like all forms of matter, is subject to the laws of physics. Here, we have provided qualitative arguments showing how many properties of living systems have their origins in fundamental physical laws. Our results suggest that despite the incredible complexity and creativity of evolution, many energetic and kinetic properties of self-replicating organisms can be understood using simple physical arguments from quantum mechanics and thermodynamics. Since the laws of physics are expected to be the same on all planets, we expect that the estimates we provide here should be equally applicable to life on other planets, if its exists. Our expressions also make a strong prediction about the variability of the three quantities we consider (growth yield, minimum doubling time and power consumption in dormancy). Namely, we expect the growth yield to be the most constrained and least variable property of chemical self-replicators across terrestrial and extraterrestrial life. The underlying reason for this is that in contrast to our expressions for the minimum doubling time and power consumption, the growth yield does not have an Arrhenius dependence and hence is not exponentially sensitivity to small changes in activation energies. In the future, it will be interesting to see if such qualitative arguments can be developed further to constrain the physical properties of chemistry-based self-replicators and aid in developing strategies for the search for extraterrestrial life.

	

\newpage



\clearpage 

%
\bibliography{PhysicalConstants.bib} 
\bibliographystyle{sciencemag}

%
%
%
%
%
%


\section*{Acknowledgments}
We would like to thank the Mehta and Kondev groups for many useful discussions. We would also like to thank Rob Phillips, Andrew Murray, and Richard Adler for comments on the manuscript. We would also like to acknowledge our colleagues in the CZI Theory group for useful comments on an early presentation of this work.
\paragraph*{Funding:}
PM was funded by the NIH NIGMS R35GM119461 and a Chan-Zuckerberg Investigator grant (to PM). JK was funded by a Simon's Investigator in Physics Fellowship and Chan-Zuckerberg Investigator grant (to JK).
\paragraph*{Author contributions:}
\section*{Supplementary materials}
Materials and Methods\\
Tables S1 to S2\\


\newpage


\renewcommand{\thefigure}{S\arabic{figure}}
\renewcommand{\thetable}{S\arabic{table}}
\renewcommand{\theequation}{S\arabic{equation}}
\renewcommand{\thepage}{S\arabic{page}}
\setcounter{figure}{0}
\setcounter{table}{0}
\setcounter{equation}{0}
\setcounter{page}{1} 


\begin{center}
\section*{Supplementary Materials for\\ \scititle}

Pankaj Mehta$^{\ast}$,
Jane Kondev$^\dagger$\\ 
\small$^\ast$ Email: pankajm@bu.edu\\
\small$^\dagger$Email: kondev@brandeis.edu
\end{center}

\tableofcontents
\listoftables

\newpage

\section{Materials and Methods}
Here we give derivations of the results in the main text. 

\subsection{Rydberg energy and Bohr radius}
Here we reproduce Weisskopf's argument for computing the Rydberg energy $Ry$ and the Bohr radius $a_0$. Consider an electron in a hydrogen atom. We would like to determine the typical distance of this electron and its typical energy $E$. The energy of the electron at a distance $r$ from the nucleus and momentum $p$ is given by 
\be
E= -{1 \over 4 \pi \epsilon_0} {e^2 \over r} + {p^2 \over 2  m_e },
\ee
where $\epsilon_0$ is the vacuum electric permittivity, $e$ is the electron charge, and $m_e$ is the mass of the electron. We can approximate the expected momentum in terms of the distance using de Broglie's relation, $p=\hbar/r$. Substituting this into the expression above yields the energy of the electron as a function of distance
\be
E= -{1 \over 4 \pi \epsilon_0} {e^2 \over r} + {\hbar^2 \over 2 m_e r^2}.
\label{SIE:E(r)}
\ee
We can now determine the typical distance $a_0$ between electron and the nucleus as the distance that minimizes this energy (i.e. ${\partial E \over \partial r}\Big|_{a_0}=0$). 

Namely, we have that
\be
{\partial E \over \partial r}\Big|_{a_0}=-{\hbar^2 \over  m_e a_0^3}+{1 \over 4 \pi \epsilon_0} {e^2 \over a_0^2}=0
\ee
Solving for $a_0$ gives
\be
a_0={4 \pi  \hbar^2 \epsilon_0 \over m_e e^2}= {\hbar \over m_e \alpha c},
\label{SIE:a0}
\ee
where we have introduced the fine structure constant $\alpha \approx 1/137$, the speed of light $c$, and used 
\be
\alpha \hbar c = {e^2 \over 4\pi \epsilon_0}.
\ee
This length scale $a_0$ is called the Bohr radius and has a numerical value of $a_0 \approx 0.53 \mathring {\mathrm A}  \approx 5.3 \times 10^{-11}$ m.
The typical energy scale, the Rydberg energy $Ry$, is given by
\be
Ry=-E(a_0)=  {m_e e^4 \over 2 \hbar^2 (4 \pi \epsilon_0)^2} ={e^2 \over 8 \pi \epsilon_0 a_0}= {1 \over 2}{\alpha^2 m_e c^2}
\label{SIE:Rydef}
\ee
and has a numerical value of $Ry \approx 13.6$ eV $\approx 2.18 \times 10^{-18}$ J. These quantities are the fundamental length and energy scales of atomic physics that we make use of in the main text and below.

\subsection{Bond energy from bond length}

We would like to also estimate the typical energy scale associated with a chemical bond. To do so, we introduce a numerical factor -- or as Weisskopf calls it a ``fudge factor'' -- that relates bonds lengths $l_{bond}$ to the Bohr radius
\be
l_{bond}=f_{bond} a_0.
\ee
For many commonly found bonds in organic molecules,  $f_{bond} \approx 3$. The typical energy $E_{bond}$ associated with an electron confined to a distance $l_{bond}$ is just given by substituting $l_{bond}$ into Eq.~\ref{SIE:E(r)}. By explicit calculation or by comparing the result with the third term in Eq.~\ref{SIE:Rydef}, we have
\be
E_{bond} \approx {Ry \over f_{bond}}.
\label{SIE:Ebond}
\ee
Numerically, this corresponds to a bond energy of about $4.5$ eV or $430$ kJ/mol. This should be compared to the measured energies of C-H bonds of $413$ kJ/mol or C-C bond
of $348$ kJ/mol.

\subsection{Calculation of mass yield of a self-replicator}

We would like to calculate $Y$, the mass yield per unit of energy consumed by a self-replicator. We assume that the the primary source of energy consumption is the formation of chemical bonds in order to self replicate organic matter. In other words, energy consumption is dominated by anabolism rather than catabolism. If an organism of dry mass $M$ requires the formation of $N_{bond}$ new bonds, then the mass yield is just
\be
Y= {M \over N_{bond} E_{bond}}.
\ee
It is reasonable to assume that the number of bonds that have to be formed during self-replication is proportional to number of  atoms:  $N_{bond} \approx  b N$, where $b$ is the proportionality constant which is order $1$. It can be be smaller if the replicator is getting large number of building blocks directly from the environment (e.g. in rich media where all amino acids are supplied we expect $b \sim 1/10$). If we denote the average mass number of an atom in the organism by $A_{bio}$, then we also have the relation
\be
N_{bond} \approx b N \approx { b M \over A_{bio} m_p},
\label{SIE:Nbond}
\ee
where $m_p$ is the mass of a proton. Substituting this in the equation above, we find that the total mass cancels and that
\begin{align}
Y &\approx {A_{bio} m_p \over b E_{bond}}=\left({A_{bio} f_{bond} \over b}\right){m_p \over Ry} 
=\left({A_{bio} f_{bond} \over b}\right){2 m_p \over m_e c^2 \alpha^2}
\end{align}
where we have used Eqs. \ref{SIE:Rydef}	and ~\ref{SIE:Ebond}. Numerically, substituting we have that
\be
Y \approx  {A_{bio} f_{bond} \over b} 8 \times 10^{-7} g/J
\ee

\subsection{Berg kinematic viscosity}

As pointed out by Ed Purcell, a wide variety of liquids seem to exhibit a similar minimum kinematic viscosity $\nu_{min}$ of around $10^{-7}$ m$^2$/s \cite{purcell2014life}. Recently, Trachenko and Brazhkin provided a simple dimensional analysis argument relating this number to fundamental physical constants \cite{trachenko2021quantum}. The basis of this argument is the observation that kinematic viscosity has its origin in quantum mechanical forces. A simple argument makes this clear. The kinematic viscosity has
units of diffusion ( $\nu_{min} \sim [length]^2/[time]$ ) and measures the ``diffusion of momentum'' in a fluid.

At high temperatures when the substance is a gas, the diffusion of momentum is controlled by thermal scattering between particles. In this regime, increasing the temperature also increases the viscosity since this increases the velocity and scattering length. In contrast,  at low temperatures when substances are cooled, the viscosity increases when the temperature is decreased (the opposite of the gaseous phase). The microscopic origin of this is that particle become kinetically trapped and rearrangements of particles is dominated by rare rearrangements due to thermal activation, a process that is empirically well modeled by the  Arrhenius Law. For this reason, the minimum of the kinematic viscosity of a liquid occurs when particles are dense enough for repulsive effects originating in quantum mechanics to be relevant but not so crowded that they get trapped kinetically. 

In this regime, the diffusion of momentum is governed by atomic length scales and  frequencies. There is only one natural length scale, the Bohr radius $a_0$ (Eq.~\ref{SIE:a0}). To calculate the kinematic viscosity, we also need to find a characteristic frequency $\omega_{D}$ governing particle interactions. This frequency should clearly depend on the mass of the fluid molecules and can be calculate by modeling particle interactions as springs. The only energy scale in the problem is the Rydberg energy (Eq.~\ref{SIE:Rydef}). Using the relation between energy and displacement for a harmonic spring $E={1\over 2}Kx^2$, we can calculate a characteristic ``spring constant'' for the problem as $ K \approx 2Ry/a_0^2$ and a characteristic angular frequency,
\be
\omega_D =\sqrt{K/m} =\sqrt{2Ry \over m a_0^2}.
\ee
Then, the typical viscosity in this regime follows from dimensional analysis and can be written as
\be
\nu_{min} \sim {\omega_D \over 2 \pi} a_0^2 \sim {1 \over 2 \pi}  \sqrt{2Ry a_0^2  \over m} \sim {1 \over 2 \pi}{\hbar \over \sqrt{m m_e}},
\ee
where in the last line we have used Eqs.~\ref{SIE:a0} and ~\ref{SIE:Rydef}. It will be helpful to express the molecular mass in terms of the proton mass as $m=A_{f} m_p$. With this substitution we have
that 
\be
\nu_{min}=  A_{f}^{-{1 \over 2}} \nu_{Berg}
\label{SIE:numin}
\ee
where we have defined the ``Berg kinematic viscosity''
\be
\nu_{Berg} \sim  {\hbar \over 2 \pi \sqrt{m_p m_e}}.
\label{SIE:nuberg}
\ee
Numerically, this gives $\nu_{Berg} \approx 4 \times 10^{-7}$ m$^2$/s. This is very close to observed minimum viscosity for a wide variety of liquids \cite{trachenko2021quantum}. For example, the kinematic viscosity of water at 100C is about $3 \times 10^{-7}$ m$^2$/s.

\subsection{Maximum rate constant for chemical reactions}

We are interested in rate at which chemical reactions happen. Following Berg and Purcell \cite{berg1977physics}, we will assume that reactions are diffusion limited. In this case, the rate $k_{on}$  at which a substrate at a concentration $c$ encounters a molecule of size $R$ is given by
\be
k_{on}=4 \pi D c R,
\label{SIE:kon1}
\ee
where $D$ is the diffusion constant of the substrate. The diffusion constant can be expressed in terms of the temperature, viscosity $\eta$, and the Stokes radius of the substrate particle $r$ using the Stokes-Einstein-Sutherland equation which holds at low-Reynolds number:
\be
D= {k_B T \over 6 \pi \eta r}
\ee
We can relate the viscosity to the kinematic viscosity by the density of the fluid $\rho_f$,
\be
\eta=\rho_f \nu= A_{f} m_p c_f \nu,
\label{SIE:etanu}
\ee
where in the second equality we have used the fact the density $\rho_f$ can be written a the fluid concentration $c_f$ times the mass of the fluid molecules $A_{f}m_p$, which we express in units of the proton mass. Substituting the last two equations into Eq.~\ref{SIE:kon1} gives
\be
k_{on}= {2 k_B T c R \over 3 \eta r}= {2 \over 3} A_{f}^{-1} \left({c \over c_f}\right)\left({R \over r}\right){k_B T  \over m_p \nu}
\ee
An upper bound on the \emph{fastest} possible on rate, $k_{on}^{max}$, at a temperature $T$ can be found by replacing $\nu$ by the \emph{minimum} kinematic viscosity $\nu_{min}$ ~(\ref{SIE:numin}). This gives an upper bound for the on rate of the form
\begin{align}
k_{on}^{max}&={2 \over 3} A_{f}^{-1} \left({c \over c_f}\right)\left({R \over r}\right){k_B T  \over m_p \nu_{min}} \nonumber \\
&={2 \over 3} A_{f}^{-{1 \over 2}} \left({c \over c_f}\right)\left({R \over r}\right){k_B T  \over m_p \nu_{Berg}} \nonumber \\
&= A_{f}^{-{1 \over 2}} \left({c \over c_f}\right) \left({R \over r}\right)\left[ \left( { m_e \over m_p}\right)^{1 \over 2} {4 \pi k_B T    \over 3 \hbar}\right]
\label{SIE:konmax}
\end{align}
The term in the bracket sets a natural scale for chemical kinetics. For a temperature $T=300$K, this takes a numerical value of 
\be
k_{on}^{max} = A_{f}^{-{1 \over 2}} \left({c \over c_f}\right) \left({R \over r}\right) 4  \times 10^{12} \mathrm{s}^{-1}.
\ee
 In other words, the characteristic time unit of chemistry is picoseconds.
 
 For cells, $R$ and $r$ are both set by the molecular scale and hence $R/r \sim 1$. For water, $A_{f}=18$, $c_f=55M$. Finally, the highest concentrations found inside cells are in the 100mM range (for example, glutamate). Thus, for
 cells, we can further estimate
 \be
 k_{on}^{max} \approx 2  \times 10^{9}\,\mathrm{s}^{-1},
 \ee 
as an upper bound on the maximum collision frequency in cells.

\subsection{Minimum barrier height for bond formation}

In the transition-state theory of chemical kinetics, the rate of a chemical reaction takes the simple form \be
k=k_{on} e^{-\Delta E \over k_B T},
\label{SEI:k}
\ee
where $k_{on}$ is an attempt rate that governs how often molecules collide (see Eq.~\ref{SIE:konmax})
and $\Delta E$ is an energetic activation barrier equal to the difference in energies between the initial state and the transition state. For this reason, a central role is played by $\Delta E$.

We will be interested in calculating $k^{max}$ -- the fastest rate at which a chemical reaction can proceed. For this reason, we are interested in calculating the minimum activation energy of the transition state, $\Delta E^{min}$. In order to model this, we assume that we can think of a bond as a spring and associate $\Delta E^{min}$ with the energy cost of stretched bond in the transition state. 

Denote the rest length of the bond is $l_{bond}$. Then the corresponding spring constant can be calculated from the bond energy: $K_{bond}= 2 E_{bond}/l_{bond}^2$. During a chemical reaction, we assume that in the transition state a bond stretches from its rest length $l_{bond}$ to a length $l_{bond} +\Delta l = l_{bond}(1 +\delta)$, where we have defined the dimensionless quantity $\delta= \Delta l/l_{bond}$. The energy cost of this stretching is
\be
\Delta E^{min} = {1 \over 2}K_{bond} (\Delta l)^2 =  E_{bond}\left( \Delta l \over l_{bond}\right)^2= E_{bond} \delta^2.
\ee
Substituting Eq.~\ref{SIE:Ebond} gives
\be
\Delta E^{min} = \left( {\delta^2 \over f_{bond}}\right) Ry =  \left({\delta^2 \over f_{bond}} \right) {\alpha^2 m_e c^2 \over 2}  
\label{SEI:Emin}
\ee
where $f_{bond}$ expresses the bond length in terms of the Bohr radius: $l_{bond}=f_{bond}a_0$. For the choice $f_{bond} =3$ and $\delta =0.3-0.5$ -- a reasonable choice for a wide variety of reactions, we get numerical values of 
\be
\Delta E^{min} \approx 0.4-1.1 \mathrm{eV}.
\label{SEI:Emin-numerical}
\ee

At room temperature $T=300K$ ($k_B T \approx 0.026$eV), this corresponds to an Arrhenius factor
 \be
 e^{\Delta E \over k_B T}  \approx e^{15} -e^{40} \approx 1 \times 10^{6} - 1 \times 10^{17}
 \ee
Notice, that the square dependence of $\Delta E$ on $\delta^2$ makes kinetic rates extremely sensitive to this exact number since the Arrhenius factor is exponential in $\Delta E$. Since we are interested in the minimum vale of $\Delta E^{min}$ we will assume that $\delta$ must be greater that $0.3$ and hence involve at least a thirty percent stretch in bond lengths. For peptide bonds in a ribosome, $\Delta E \approx 0.6-0.8$\,eV \cite{Wallin2010_peptidyl_transfer_TS}.

\subsection{Minimum doubling time}

Before a self-replicator can divide, it must first double its size. This takes both time and energy. There are two possibilities for what sets the minimum doubling time: (i) self-replication is kinetically limited and (ii) self-replication is energy-limited. When self-replication is kinetically limited, the minimum doubling time is set by the time it takes to reproduce molecules (anabolism). In contrast, when the self-replicator is energy limited, the rate limiting step for self-replication is extracting energy from the environment (catabolism). \\

\noindent \underline{{Kinetically limited self-replicators:}}\\\\
We now calculate the minimum doubling time for a self-replicator that is kinetically limited. The fastest self-replicator is an organism where each molecule replicates itself. If a molecule on average has $N$ bonds that need to be synthesized, then the minimum doubling time can be written in terms of the maximum rate of a chemical reaction $k^{max}$ as 
\be
T_{min}= {N \over k^{max}}= {N\over k_0^{max}}e^{\Delta E \over k_B T},
\label{SEI:tau-exp}
\ee
where we have used Eq.~\ref{SIE:konmax} for $k^{max}$. 

Let us evaluate this assuming that protein synthesis is the limiting kinetic step. Assuming every protein translates itself, $N$ will equal the typical number of peptide bonds in all ribosomal proteins, $N \sim 7400$.  Using Eqs~\ref{SIE:konmax},
at room temperature we have
\be
 k_0^{max} \sim A_{f}^{-{1 \over 2}} \left({c \over c_f}\right) \left({R \over r}\right)4  \times 10^{12} \mathrm{s}^{-1}.
 \ee
 For chemical kinetics, the target and substrate radii $R$ and $r$ are the same order of magnitude. The typical concentration $c$ of amino acids is 100-300mM. For water, $A_{f} \approx 18$ and $c_f \approx 55M$. With these substitutions we have
 \be
 k_0^{max} \approx 2 \times 10^{9} \,\mathrm{s}^{-1}
 \ee
 The exponential factor can be estimated from Eq.~\ref{SEI:Emin-numerical}. We see that $\Delta E$ equal ranges from $0.4\,\mathrm{eV} -1.1 \, \mathrm{eV}$ so that $
 e^{\Delta E \over k_B T}  \approx e^{15} -e^{40} \approx 1 \times 10^{6} - 1 \times 10^{17}
$ (see Eq.~\ref{SEI:Emin-numerical}).
 Substituting these expressions into  Eq~\ref{SEI:tau-exp} we have
 \be
T_{min} \approx 10\,\mathrm{s} -10^{12} \,\mathrm{s}
 \ee
 The large range is due to the exponential dependence of $T$ on the transition state energy $\Delta E$.

\noindent \underline{{Energetically limited self-replicators:}}\\\\

We now consider the case where self-replication is limited by catabolism-- the ability to harvest energy from the environment. This is the case for extreme environments such as ammonia oxidizers in the oligotrophic ocean or lithoautotrophs deep in the ocean crust. In this case, growth is limited by the ability of to extract energy from the environment. 

We will assume that the bacteria are absorbing spheres of typical size $R$ and that the electron donors are found at  concentration $c$, which for ammonia-oxidizing archaea is typically in the nM range. We assume that bacteria can extract an energy $E_{redox}$ per molecule. This is set by the chemistry of redox reactions and is typically of order $0.1-1 eV$, or in terms of the Rydberg energy between $Ry/100-Ry/10$. The rate at which bacteria can extract energy is then limited by the diffusion limited flux incident on the cell (Eq.~\ref{SIE:konmax}):
\be
k_{on}^{max}= A_{f}^{-{1 \over 2}} \left({c \over c_f}\right) \left({R \over r}\right)\left[ \left( { m_e \over m_p}\right)^{1 \over 2} {4 \pi k_B T    \over 3 \hbar}\right],
\label{SEI:kenergy}
\ee
where $A_{f}$ and $c_f$ are the atomic number and concentration of the fluid, $R$ is the size of the absorber, in this case assumed to be the entire surface area of the cell, and $r$ is the size of solute molecules. The total power consumption of the bacterial cell is just $P^{max}=k_{on}^{max} E_{redox}$. 

The total energy needed to reproduce is the $E_{tot}=N E_{bond}$, where $N$ is the number of bonds. We assume, as in Eq.~\ref{SIE:Nbond} that $N_{bond} \approx b N \approx { b M \over A_{bio} m_p}$, where $b \approx 1-4$ is the number of bonds per atom, $A_{bio}$ is the atomic number of biomass, and $M \approx 10^{-12}$ g is the mass of a cell. The total time that a self-replicator will take is just
\be
T_{min} = {E_{tot} \over P^{max}} = \left({b M \over A_{bio} m_p k_{on}^{max}}\right){E_{bond} \over E_{redox}}
\ee
Substituting Eq.~\ref{SEI:kenergy}, for a self-replicator with mass $M$ we have
\begin{align}
T_{min} &= b A_{bio}^{-1} A_{f}^{{1 \over 2}} \left({c \over c_f}\right)^{-1} \left({R \over r}\right)^{-1}M  \left({3   \hbar \over 4 \pi k_B T \sqrt{m_p m_e}}\right){E_{bond} \over E_{redox}}\\
&= b A_{bio}^{-1} A_{f}^{{1 \over 2}} \left({c \over c_f}\right)^{-1} \left({R \over r}\right)^{-1} M \left({3 \nu_{berg} \over 2  k_B T }\right){E_{bond} \over E_{redox}}
\label{SIE:energy-limited}
\end{align}

For ammonia-oxidizing archaea, the limiting electron donor is ammonia, dissolved in water at a concentration of about $1$ nM. For these bacteria, the relevant fluid is water so that $A_{f} \approx 18$ and $c_f \approx 55 M$, $R \approx 1 \mu$ m (the size of the cell), $r \approx 1$ nm is the size of solute molecules, $E_{redox} \approx 0.77$ eV,  $M \approx 10^{-12}g$, $b \approx 1$, and $A_{bio} \approx 13$. Plugging in these numbers we have
\be
T_{min}  \approx 2 \times 10^7 \,\mathrm{s},
\ee
or approximately $6$ months. 

Lithoautotroph bacteria in the deep sea crust are thought to have doubling times in the range of  years or longer ($> 3 \times 10^{7}$ s), longer than ammonia oxidizers. This can be understood by noting these bacteria oxidize $H_2$ which is thought to be found at sub-nanomolar range $c \approx 10^{-10}-10^{-12}$ M. Substituting into Eq~\ref{SIE:energy-limited}, one finds $ T_{min}  \approx 2 \times 10^8- 2 \times 10^{10}\,\mathrm{s}$, consistent with experiments  \cite{ lloyd2020evidence}.

\subsection{Membrane voltage, current, and power}
We now consider a cell with a membrane and the minimal energetic cost of maintaining a voltage across the membrane. We assume that the concentration of charged ions inside the cell, $c_{i}$, differs substantially from the concentration of ions outside a cell $c_{o}$. For example, in {\em E. coli}, K$^+$ ions are typically at a concentration of $300$~mM in the cytoplasm of the cell, while in the environment (LB medium) $10$~mM is the norm.  Then, from standard thermodynamic arguments we know that the free energy difference for ions inside and outside the membrane due this concentration difference is 
\be
\Delta G= k_B T \ln{\left({c_{o} \over c_{i}}\right)} ,
\ee 
As the ions cross the membrane down the concentration gradient an electrical potential difference builds up. At equilibrium, the free energy difference $ \Delta G$ is counteracted by electrical potential energy. If we denote the membrane voltage by $V_m$, then energy balance gives us the Nernst equation, with $z$ the charge of the ion,
\be
V_m = {\Delta G \over z e} = {k_B T \over z e}\ln{ \left({c_{o} \over c_{i}}\right)}.
\label{SIE:Vm}
\ee
The prefactor ${k_B T \over  e}$ is often called the thermal voltage and is equal to approximately $26$ mV at room temperature ($T=298$ K). This sets the scale of the voltage across the membrane of all cells. For the case of an {\em E. coli} cell, the equilibrium voltage for potassium ions is estimated, based on Eq~\ref{SIE:Vm} to be $ V_m = 25.7 \times \ln(10 /300)$mV $ = -90$mV; the range of membrane potentials for  {\em E. coli} under physiological conditions is $-80$ to $-140$mV.  

We can also calculate the conductance of the membrane. To do so, we note that charged ions with charge $z$ move across the membrane primarily through ion channels, which are transmembrane proteins that provide safe passage for ions by creating a water-like electrical environment within the channel formed by the protein. Whether an ion channel is open or closed is tightly regulated by the cell which provides the cell with the means of controlling the electrical properties of its cytoplasm and the membrane voltage. 

Even in the absence of ion channels a lipid membrane eventually lets ions pass through. Thermally generated pores in the membrane provide passage that is not blocked by the large energy barrier imposed by the oily nature of the membrane interior. For determining the maintenance energy we primarily focus on ions passing through thermally generated pores as a mechanism for loss of membrane voltage. To get at the power dissipated by ions passing through a pore we first estimate the conductance of such a pore and then the average number of such pores in a membrane of a cell.  

Inside a membrane pore, the membrane potential $V_m$ gives rise to an electric field $E=V_m/d_m$, where $d_m$ is the membrane thickness. This electric field exerts a force on the particles which is counteracted by frictional drag. We can write this force balance equation as 
\be
z e E = \gamma v_{i},
\ee
where $\gamma$ is a viscous drag coefficient and $v_{i}$ is the average ion velocity. At low Reynolds number, the drag coefficient $\gamma$ can be related to the viscosity through the Stokes relation $\gamma=6 \pi \eta r$, where $r$ is the Stokes radius of the diffusing ions. Using Eq.~\ref{SIE:etanu} relating the $\eta$ to the kinematic viscosity $\nu$ and $E=V_m/d_m$, the velocity of the ions can be expressed as 
\be
v_{i} ={z \over A_{f} c_{f}  r d_m} \frac{ e V_m}{6 \pi  m_p \nu},
\ee
where $A_f$ and $c_f$ are the atomic number and concentration of the fluid respectively.

If we denote the cross-sections surface area occupied by the pore by $S_{p}$, the ion concentration by $c$, the maximum total current across the membrane $I^{max}$ is given by
\begin{align}
I^{max} &=ze c S_{p} v_{i}^{max} \nonumber \\
&= {z^2 \over A_{f}}\left({c \over c_{f}}\right)\left({S_{p} \over d_m r}\right) \frac{ e^2 V_m}{6 \pi  m_p \nu_{min}} \nonumber \\
&= {z^2 \over A_{f}^{1 \over 2}}\left({c \over c_{f}}\right)\left({S_{p} \over d_m r}\right) \frac{ e^2 V_m}{6 \pi  m_p \nu_{Berg}} \nonumber \\
&= {z \over A_{f}^{1 \over 2}}\left({c \over c_{f}}\right)\left({S_{p} \over d_m r}\right)\left[\ln{ \left({c_{o} \over c_{i}}\right)}\right] \frac{ e k_B T }{3\hbar} \sqrt{m_e \over m_p},
\end{align}
where in going to the second line we have used Eqs~\ref{SIE:numin} for the minimum viscosity and ~\ref{SIE:Vm} for the membrane potential. The maximum conductance of membrane pore $G^{max}_p$ also follows from the expression above by noting that
\be
I^{max}=G^{max}_p V_{m},
\ee
which yields
\begin{align}
G^{max}_p &= {z^2 \over A_{f}}\left({c \over c_{f}}\right)\left({S_{p} \over d_m r}\right) \frac{ e^2 }{6 \pi  m_p \nu_{min}} \nonumber \\
&={z^2 \over A_{f}^{1 \over 2}}\left({c \over c_{f}}\right)\left({S_{p} \over d_m r}\right) \frac{ e^2 }{3 \hbar}\sqrt{m_e \over m_p} 
\end{align}
To estimate the maximum conductance of a membrane pore we take  $c_f =55$~Mol for water and $c \approx 300$~mM. The membrane thickness is typically $d_m=4$nm while we take the pore diameter to be $6$\AA, i.e., the size of a hydrated K$^+$ ion. This gives $S_{p}=30$\AA$^2$ for the pore cross section area. The radius of a hydrated ion $r$ is $ r \approx 3 \times 10^{-10}$ m, and $\ln{ \left({c_{o} \over c_{i}}\right)} \approx 3$, while for water $A_f \approx 18$. Plugging in these numbers gives the maximum current and conductance of a membrane pore $G^{max}_p \approx  600 \times 10^{-12}$ S, $I^{max}= 50 \times 10^{-12}$ A. For comparison, the typical current through open ion channels are in the few pico-Ampere range while their conductances are in the $10-100$~pS range, both about an order of magnitude below our estimates. Note that had we used the kinematic viscosity of water, which at room temperature is $10^{-6} m^2/s$ or about an order of magnitude larger than the (universal) minimum viscosity, both estimates of the conductance and the current would be an order of magnitude smaller and therefore close to the measured values. 

Using the estimates for the maximum current and membrane voltage we can calculate the maximum power released by ions moving across a membrane pore:
\bea
P^{max}&=&V_m I^{max} = {1 \over A_{f}^{1 \over 2}}\left({c \over c_{f}}\right)\left({S_{p} \over d_m r}\right)\left[\ln{ \left({c_{o} \over c_{i}}\right)}\right]^2 \frac{ (k_B T)^2 }{3\hbar} \sqrt{m_e \over m_p}\\
&=& {1 \over 3} A_{f}^{-{1 \over 2}}\left({c \over c_{f}}\right)\left({S_{p} \over d_m r}\right)\left[\ln{ \left({c_{o} \over c_{i}}\right)}\right]^2 \frac{ k_B T }{ \tau_{min}(T)} 
\label{SIE:Pmax}
\eea

Equation~\ref{SIE:Pmax} yields the estimate $P^{max} \approx 4 \times 10^{-12} W$ for the power expended by an ion current through a single membrane pore. Again, this is about an order of magnitude larger than the estimate we get if we use the actual kinematic viscosity of water instead of $\nu_{min}$.  To obtain an estimate of the maximum power expended by the cell to maintain the balance of ions inside and outside the cell, we still need to estimate the expected number of pores in the cell membrane, as each pore will require the cell to expend $P_{max}$ energy to counteract the free energy expended by the ion current. To  estimate  the average number of pores we make use of the idea that the molecular-scale energy of interaction between lipids in a membrane is of the order of 0.3eV, about an order of magnitude less than the scale of the covalent bond energy which itself was a few times less than the Rydberg energy. 

This energy scale of non-covalent bonding between molecules, follows from a simple model of a molecule as a dipole whose dipole moment is $p_d = e l_{bond}$ and assuming that the interaction between two molecules is a dipole-dipole interaction. If the typical distance between the interacting dipoles is $l_{int}$ then the estimate of the non-covalent bond energy is 
\be
E_{int}  = {1 \over 4 \pi \epsilon_0} {p_d^2 \over l_{int}^3} = \left ( {1 \over 4 \pi \epsilon_0} {e^2 \over  2 a_0} \right ) {2 a_0 l_{bond}^2 \over l_{int}^3} \ .
\ee
Using our previous estimate for $l_{bond} = f_{bond} a_0$ and a new estimate for $l_{int} = f_{int} a_0$ where $f_{int}$ is about a three times the molecular scale, or about ten times the atomic scale (i.e.,$f_{int} \approx 10$), we arrive at the estimate
\be 
E_{int} \approx  0.3 eV
\ee
which is to be compared to the range of non-covalent bond interactions for biomolecules that spans 0.03 to 0.3 eV, from the weakest London forces to the strongest hydrogen bonds \cite{lehninger2004}. The same estimate is obtained by using the areal modulus of a lipid membrane $K_A=0.3$ J/m$^2$. To create a pore of area $S_p=30$ \AA$^2$, the energy required is $1/2(0.3\, \text{J/m}^2 \times 30\, \text{\AA}^2)=0.3$ eV, the factor of $1/2$ accounting for the two layers of the lipids in the bilayer.

With this energy scale in hand we can estimate the probability that a pore opens up an lipid bilayer. We estimate that the energy needed to pull two lipids in a layer to a distance $l_{int}$ is $E_{int}$. Therefore the probability of such a separation between two lipids happening spontaneously is $p_{int} \approx e^{-E_{int}/k_BT}$. For a pore to form spontaneously across the bilayer two lipids in both layers have to separate to a distance $l_{int}$ at the same time, leading to an estimate: 
\be
p_{pore} = p_{int}^2 \approx e^{- 2 E_{int}/k_BT} \approx e^{-23.3} \approx 7 \times 10^{-11} 
\ee 
Since the number of possible pore sites, each of area $\pi l_{int}^2$, in an {\em E.coli} membrane is $S_{mem}/ \pi l_{int}^2 \approx 7 \times 10^6$, with $S_{mem} \approx 6 \mu\text{m}^2$, the total number of pores at any given time is
\be
n_{pore} = {S_{mem} \over \pi l_{int}^2} p_{pore} = 5 \times 10^{-4}
\ee
in other words the chance of a single pore opening in the membrane of an E.coli cell is about one in two thousand. This implies that the energy loss due to ions leaking through a spontaneously formed pore is 
\be
P_{dorm} \approx  5 \times 10^{-4} \times P_{max} =  2 \times 10^{-15} \mathrm{W/cell}.
\ee
This estimate goes down by an order of magnitude if we use the value for the kinematic viscosity of water, instead of $\nu_{min}$. We can also compare it to the measured maintenance cost in recent lab experiments, which are of order $10^3$~ATP/s or $10^{-16}$~W per cell \cite{ciemniecki2024mechanistic}. Similar values of the maintenance cost were estimated based on measurements of metabolic waste from a variety of bacterial cells observed in different natural environments \cite{price2004temperature}.

\newpage

%
%
%
\begin{table} 
	\centering
	\caption{\textbf{Self-replicators are constrained by physical laws governed by fundamental physical constants.}
		A list of the fundamental constants and their numerical values relevant for understanding chemistry-based self-replicating systems. Units are meters (m), seconds (s), Joules (J), electron volts (eV), Coulombs (C), kilograms (kg).}
	\label{tab:fundamentalconstants} 
	
	\begin{tabular}{lccc} 
		\\
		\hline
		Constant& Symbol & \multicolumn{2}{c}{Value}  \\
		 & &  (SI units) & (Alternative units) \\
		\hline
		Speed of light & $c$ &  $3.0 \times 10^8$ m$\cdot$s$^{-1}$ & \\
		Planck constant & $h$ & $6.6 \times 10^{-34}$ J$\cdot$s & $4.1 \times 10^{-15}$ eV$\cdot$s\\
		Elementary charge & e & $1.6 \times10^{-19}$ C & \\
		Fine structure constant & $\alpha$ & $7.3 \times 10^{-3}$ & $\approx {1 \over 137}$\\
		Mass of electron & $m_e$ & $9.1 \times 10^{-31}$ kg & \\
		Mass of proton & $m_p$ & $1.7 \times 10^{-27}$ kg & \\
		Boltzmann constant & $k_B$ & $1.4 \times 10^{-23}$ J$\cdot$K$^{-1}$ & $8.6 \times 10^{-5}$ eV$\cdot$K$^{-1}$\\
		\hline
	\end{tabular}
\end{table}

\begin{table} 
	\centering
	\caption{\textbf{The qualitative behavior of self-replicators can be expressed in terms of characteristic physical scales.}
		The fundamental constants in Table \ref{tab:fundamentalconstants}, can be combined in various ways to write
		characteristic physical scales that constrain the properties of self-replicators. $\hbar=h/2\pi$ is the reduced Planck constant and all numerical values are calculated at room temperature ($T=298K$).}
	\label{tab:physcales} 
	
	\begin{tabular}{lcccc} 
		\\
		\hline
		Physical Scale & Symbol &  Expression & \multicolumn{2}{c}{Value}  \\
		& &  & (SI units) & (Alternative units) \\
		\hline
		\rule{0pt}{25pt}Bohr radius & $a_0$ & $\frac{\hbar}{m_e c \alpha}$ & $5  \times 10^{-11}$ m  &\\
		\rule{0pt}{25pt} Rydberg energy & Ry & ${1 \over 2}m_e c^2 \alpha^2$ & $2  \times 10^{-18}$ J& $13.6$eV\\
		\rule{0pt}{25pt}Thermal energy &   & $k_B T$ &  $4  \times 10^{-21}$ J & 25.7meV \\
		\rule{0pt}{25pt}Min. kinematic viscosity (Berg Viscosity) & $\nu_{B}$ & ${1 \over 2 \pi} {\hbar \over \sqrt{m_e m_p}}$ & $4 \times 10^{-7}$ m$^2 \cdot$s$^{-1}$& \\
		\rule{0pt}{25pt}Min. time scale for chemical kinetics & $\tau_{min}(T)$ & ${\hbar \over k_B T} \sqrt{m_p \over m_e}$ &  $1 \times 10^{-12}$ s &\\
		\rule{0pt}{25pt} Chemical assembly constant & $ Y_c$  &${2 m_p \over m_e c^2 \alpha^2}$ &   $8 \times 10^{-7}$ g/J&
		\rule{0pt}{-5pt}\\
		\hline
	\end{tabular}
\end{table}

\end{document}